\documentclass{PoS}
\usepackage{graphicx}
\title{Highlight of top quark properties measurements at ATLAS/CMS}

\ShortTitle{Highlight of top quark properties}

\author{\speaker{Prolay Kumar Mal}\thanks{on behalf of ATLAS, CMS and Top Collaboration}\\
  School of Physical Sciences\\
  National Institute of Science Education and Research\\
  Bhimpur-Padanpur (Jatni), Dist. Khorda, Odisha, India 752050\\
  E-mail: \email{prolay.kumar.mal@cern.ch}}

\abstract{The top quark is the heaviest known elementary particle and plays a
special role in the dynamics of fundamental interactions. Since its discovery
at the Tevatron, several of its properties have been measured by the Tevatron
experiments (CDF and D\O). However, thanks to its unprecedentedly large
production rate at the LHC a new level of precision in these measurements has
been achieved by the LHC experiments (ATLAS and CMS). The latest
LHC measurements of the top quark mass, total decay width, top-antitop
spin correlations, and charge asymmetry are presented in this contribution.
In addition, the results from the W-boson helicity measurement are presented.
}

\FullConference{XIV International Conference on Heavy Quarks and Leptons (HQL2018)\\
		May 27- June 1, 2018\\
		Yamagata Terrsa, Yamagata, Japan}

\begin{document}

\section{Introduction}\label{sec:introduction}
Within the Standard Model (SM), the top quark is the weak-isospin
partner of the bottom quark with charge and isospin quantum numbers
Q = 2/3 and $\rm T_3$ = +1/2 respectively. Being heavy, the
top quark life-time is shorter than the hadronization time-scale
and thus it provides a unique opportunity to study the bare quark.
Furthermore, due to its large mass the top Yukawa coupling to the
Higgs boson is of the order of unity and has direct implication on
the Higgs boson mass. In addition, many beyond the Standard
Model (BSM) particles are predicted to decay through top/antitop quarks.
All these characteristics of
the top quark make it special in the SM and in many extensions thereof.
It was discovered jointly by the CDF and D\O\ experiments during
Tevatron's Run 1 (1992-95). Since then many of its properties (mass,
couplings, production cross-section, decay widths, charge
asymmetry, spin correlation, etc.) have been studied extensively during
Tevatron's Run 1 and Run 2 (2002-11).
\par

At the LHC, the top quarks are predominantly produced in pairs of
top and antitop quarks ($\rm t\bar{t}$) through strong interactions.
In addition, they can be produced through electroweak interactions
at a relatively lower production rate. Within the Standard Model,
the top quark decays into a W-boson and a bottom quark with almost 100\% branching
fraction and thus the final state signature gets driven by the decay
modes of the W-boson. In the subsequent sections, the latest LHC results
on top quark properties by the ATLAS~\cite{reference:atlascollaboration}
and CMS~\cite{reference:cmscollaboration} collaborations are presented
using the top quark sample from LHC Run 1 and Run 2 (at center of
mass energies, $\rm\sqrt{s}$ = 7, 8, and 13 TeV). The LHC Top Working Group~\cite{reference:lhctopwg}
has combined some of these results and the combined sensitivity is also
presented here whenever it is available.

\section{Top Charge Asymmetry}\label{sec:asymmetry}
\begin{figure*}
\begin{center}
\begin{tabular}{cc}
\includegraphics[width=0.45\textwidth]{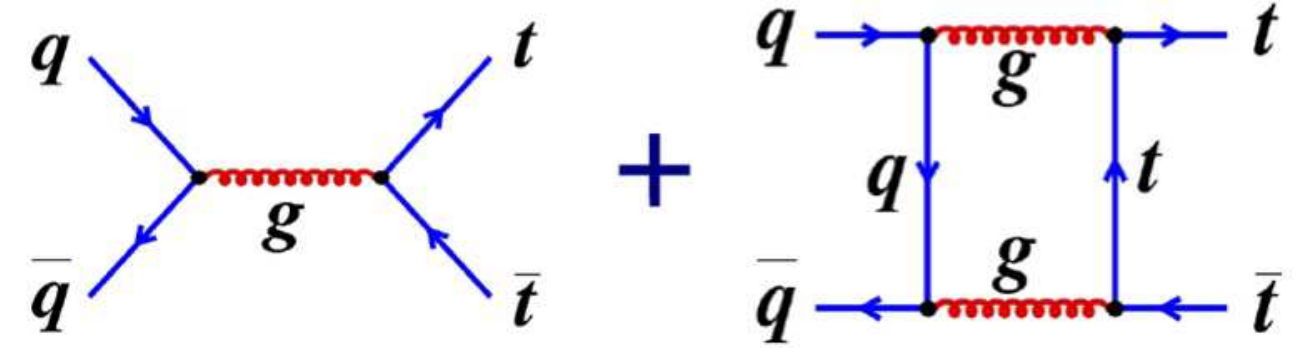}&\includegraphics[width=0.45\textwidth]{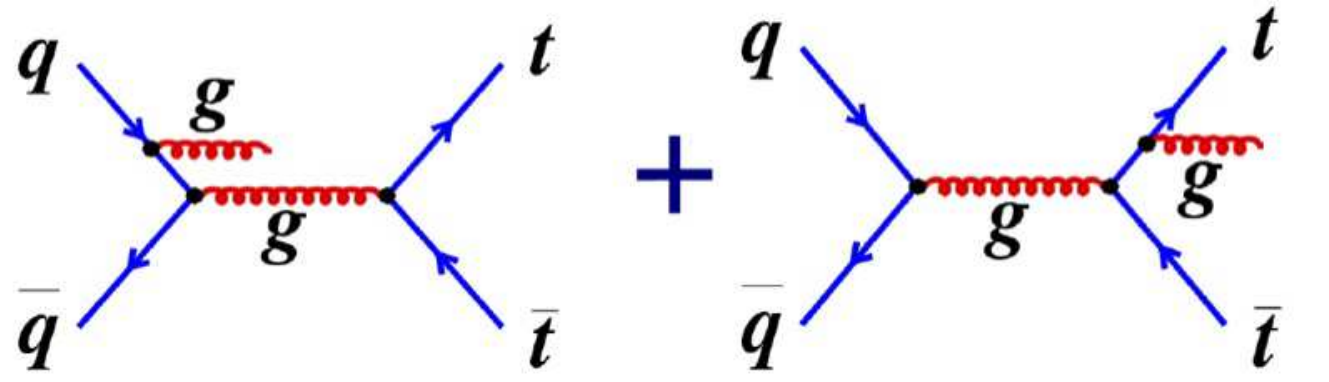}\\
(a) & (b)
\end{tabular}
\end{center}
\caption{Feynman diagrams for (a) positive $t\bar{t}$ asymmetry from interference of
the LO and the box diagrams, (b) negative $t\bar{t}$ asymmetry from interference of
the Initial State Radiation (ISR) and the Final State Radiation (FSR).}
\label{fig:asymmetry}
\end{figure*}

Leading Order (LO) Quantum Chromodynamics (QCD) predicts a completely
symmetric $\rm t\bar{t}$ pair production at hadron collider, i.e., the
top (antitop) quarks do not have any preferential direction with respect
to the initial quarks (anti-quarks). The contributions from
gluon-gluon fusion processes, i.e., $\rm gg\rightarrow t\bar{t}$
are also completely symmetric and thus dilute the overall
$\rm t\bar{t}$ charge asymmetry. However, at next-to-leading order
(NLO) QCD~\cite{reference:asymmetry}, an asymmetry originates from the
interference of different Feynman diagrams as shown in Fig.~\ref{fig:asymmetry}.
The $\rm t\bar{t}$ charge asymmetry can further get enhanced through contributions
from the BSM physics, e.g., exchange of axigluons, heavy Z particles, or colored
Kaluza-Klein gluon excitations~\cite{reference:asymmetry_bsm}.
\par
\begin{figure*}
\begin{center}
\begin{tabular}{cc}
\includegraphics[width=0.5\textwidth]{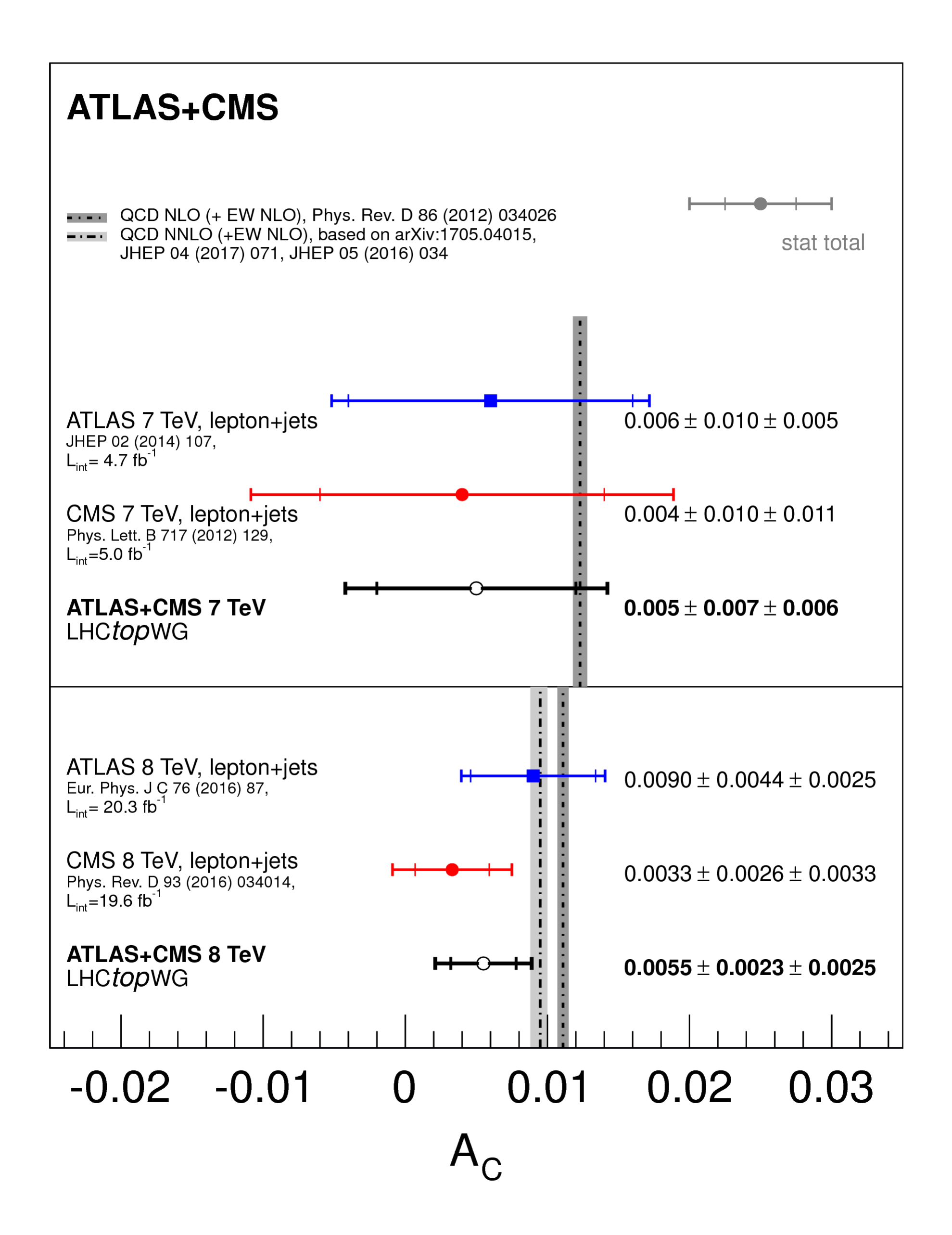}&\includegraphics[width=0.5\textwidth]{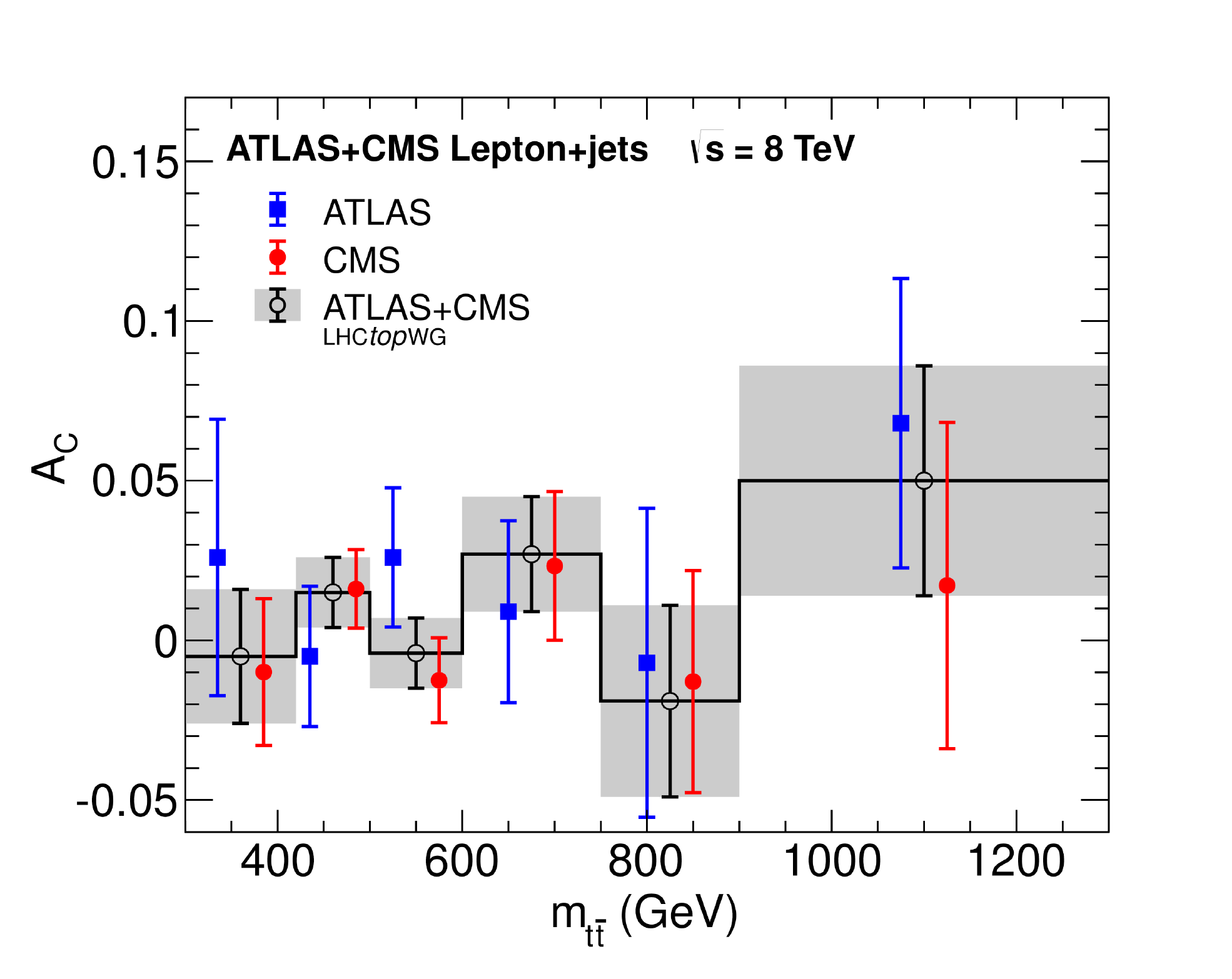}\\
(a) & (b)
\end{tabular}
\end{center}
\caption{\cite{reference:asymmetry_results_lhctop} (a) Summary of inclusive $\rm A_C^{t\bar{t}}$ measurements by ATLAS and CMS, and the LHC combination
at $\rm\sqrt{s}$ = 7 and 8 TeV, in comparison with the theoretical predictions at NLO and NNLO; (b)
$\rm A_C^{t\bar{t}}$ vs $\rm m_{t\bar{t}}$ at $\rm\sqrt{s}$ = 8 TeV from ATLAS, CMS and the LHC combination.}
\label{fig:asymmetry_results_lhctop}
\end{figure*}

In recent times, the legacy measurements by the Tevatron experiments (using semileptonic
and dileptonic $\rm t\bar{t}$ events) show some deviations~\cite{reference:asymmetry_tevatron}
from the SM QCD predictions in terms of $\rm t\bar{t}$ forward-backward asymmetry ($\rm A_{FB}$).
However, at the LHC, being a proton-proton collider with much higher center-of-mass energy, the
$\rm A_{FB}$ becomes undefined and a related charge asymmetry ($\rm A_C$) is considered using
the absolute rapidity differences between the top and antitop quarks,
\begin{eqnarray*}
\rm A_C =\frac{N(\Delta |y| > 0) - N(\Delta |y| < 0) }{N(\Delta |y| > 0) + N(\Delta |y| < 0)},
\end{eqnarray*}
where $\rm\Delta |y|= |y_t| -|y_{\bar{t}}|$ with $\rm y$ being the rapidities of top and antitop quarks.
ATLAS has performed the $\rm A_C$ measurements~\cite{reference:asymmetry_atlas_dilepton} in the dilepton
channel using the Run 1 dataset ($\rm\sqrt{s}$ = 8 TeV), where inclusive $\rm t\bar{t}$ and leptonic
asymmetries respectively result in $\rm A_C^{t\bar{t}} =0.021 \pm 0.016$ and $\rm A_C^{l^+l^-} = 0.008 \pm 0.006$.
In addition, the differential measurements are performed as functions of the invariant mass of
the $\rm t\bar{t}$ system ($\rm m_{t\bar{t}}$), the transverse momentum of the $\rm t\bar{t}$ system
($\rm p_{T,̄t\bar{t}}$),
and the absolute boost of the $\rm t\bar{t}$ system along the beam axis ($\rm \beta_{z,t\bar{t}}$).
Using the 2012 dataset, CMS observes~\cite{reference:asymmetry_cms_dilepton} the values of
$\rm 0.011\pm 0.011 (stat) \pm 0.007 (syst)$ and $\rm 0.003 \pm 0.006 (stat) \pm 0.003 (syst)$,
respectively for the $\rm t\bar{t}$ and leptonic inclusive charge asymmetries.
In the lepton+jets channel, the ATLAS measurements~\cite{reference:asymmetry_atlas_ljets}
in a fiducial region with $\rm m_{t\bar{t}}>0.75\  TeV$ and $\rm -2.0<|y_t|-|y_{\bar{t}}|<2.0$ lead to a value
of $\rm A_C^{t\bar{t}} = 0.042 \pm 0.032$. With the template techniques, the CMS measurements
\cite{reference:asymmetry_cms_ljets} in the lepton+jets channel using the $\rm 19.7\  fb^{-1}$ of the Run 1
dataset (at $\rm \sqrt{s}$ = 8 TeV) lead to the most accurate value of
$\rm A_C^{t\bar{t}} = 0.0033 \pm 0.0026(stat) \pm 0.0033(syst)$. The LHC Top Working Group
has recently combined all the Run 1 legacy results from the ATLAS and CMS
experiments using the Best Linear Unbiased Estimator (BLUE) and has reported
\cite{reference:asymmetry_results_lhctop}
the inclusive and differential measurements on $\rm A_C^{t\bar{t}}$ as shown in
Fig.~\ref{fig:asymmetry_results_lhctop}. In summary, all the $\rm t\bar{t}$ charge asymmetry
results (at inclusive and differential levels) from the ATLAS and CMS experiments using the
Run 1 legacy datasets are consistent with the SM predictions.
\section{Top Spin Correlation}
In the QCD-dominated $\rm t\bar{t}$ production process, the SM predicts only a small net
polarization for the top quarks originating from the electroweak corrections, but
the spins of the top and antitop quarks are expected to be correlated. However, the
BSM couplings~\cite{reference:spincorrelation_bsm}
of the top quark to new particles can significantly alter both the
top quark polarization and the strength of the $\rm t\bar{t}$ spin correlation.
Therefore, a precise measurement of the $\rm t\bar{t}$ spin correlation can probe the
top quark couplings to the BSM particles.
\par
In $\rm t\bar{t}$ pair produced events, the spin orientations of the top and antitop quarks
are manifested through the decay products and thus the $\rm t\bar{t}$ spin correlation can be
measured directly through their angular distributions.
In dileptonic signatures of the $\rm t\bar{t}$ events, the difference in azimuthal angle
between the charged leptons in the laboratory frame, $\rm\Delta\phi_{l^+l^-}$ is sensitive
to the $\rm t\bar{t}$ spin correlation and can be measured precisely without the full reconstruction
of $\rm t\bar{t}$ event kinematics. The reconstructed $\rm\Delta\phi_{l^+l^-}$ distributions can
then be utilized to extract a coefficient,
$\rm f_{SM}=N_{SM}^{t\bar{t}}/(N_{SM}^{t\bar{t}}+N_{uncor}^{t\bar{t}})$, to measure the degree of spin
correlation relative to the SM prediction. Here, the $\rm N_{SM}^{t\bar{t}}$ and $\rm N_{uncor}^{t\bar{t}}$
respectively represent the numbers of $\rm t\bar{t}$ events with the SM-predicted spin correlation
and without any spin correlation. A negative value of $\rm f_{SM}$ would indicate an anti-correlation
between top and antitop quark spins, while $\rm f_{SM}=1$ would indicate the SM-predicted spin correlation.
\begin{figure*}
\begin{tabular}{cc}
\includegraphics[width=0.45\textwidth]{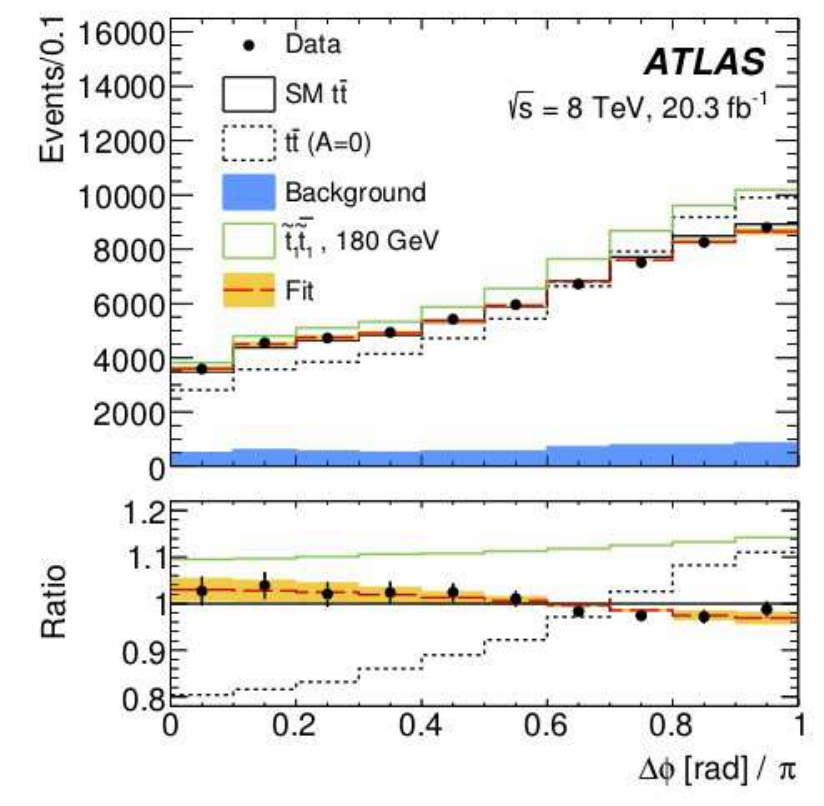}&\includegraphics[width=0.45\textwidth]{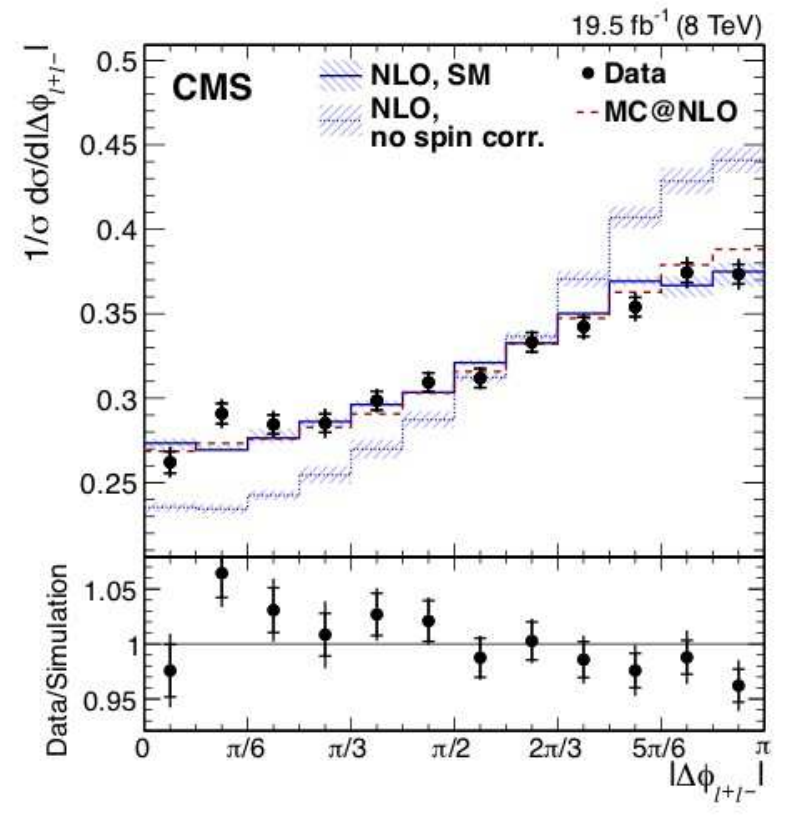}\\
(a) & (b)
\end{tabular}
\caption{\cite{reference:spincorrelation_atlas}\cite{reference:spincorrelation_cms}
(a) ATLAS $\rm\Delta\phi_{l^+l^-}$ distributions with NNLO+NNLL normalization; The NLO
predictions for stop pair-production is also shown. (b) CMS normalized differential
cross section as a function of $\rm\Delta\phi_{l^+l^-}$ at NLO predictions with (SM) and
without spin correlations; Parton-level predictions from MC@NLO is also overlaid.}
\label{fig:spincorrelation_dphill}
\end{figure*}
\par
Both ATLAS~\cite{reference:spincorrelation_atlas} and CMS~\cite{reference:spincorrelation_cms}
have measured the $\rm t\bar{t}$ spin correlation in the dilepton channels, and the corresponding
$\rm\Delta\phi_{l^+l^-}$ distributions are shown in Fig.~\ref{fig:spincorrelation_dphill}. ATLAS uses
a binned likelihood method to extract a combined ($\rm ee$, $\rm\mu\mu$, $\rm e\mu$) fit value of
$\rm f_{SM}=1.20\pm 0.05 (stat)\pm 0.13 (syst)$, which agrees with the SM prediction within two standard
deviations. The observed $\rm\Delta\phi_{l^+l^-}$ distribution is utilized to look for the top squark
pair production ($\rm\tilde{t_1}\bar{\tilde{t_1}}$) signature with
$\rm\tilde{t_1}\rightarrow t\tilde{\chi^0_1}$ decays. However, no evidence for 
$\rm\tilde{t_1}\bar{\tilde{t_1}}$ production has been found and
the range of $\rm m_t< m_{\tilde{t_1}}< 191$ GeV has been excluded at 95\% Confidence Level (CL) assuming
BR($\rm\tilde{t_1}\rightarrow t\tilde{\chi^0_1}$)=100\% and $\rm m_{\tilde{\chi^0_1}}$=1 GeV.
\par
\begin{figure*}
\begin{center}
\includegraphics[width=0.75\textwidth]{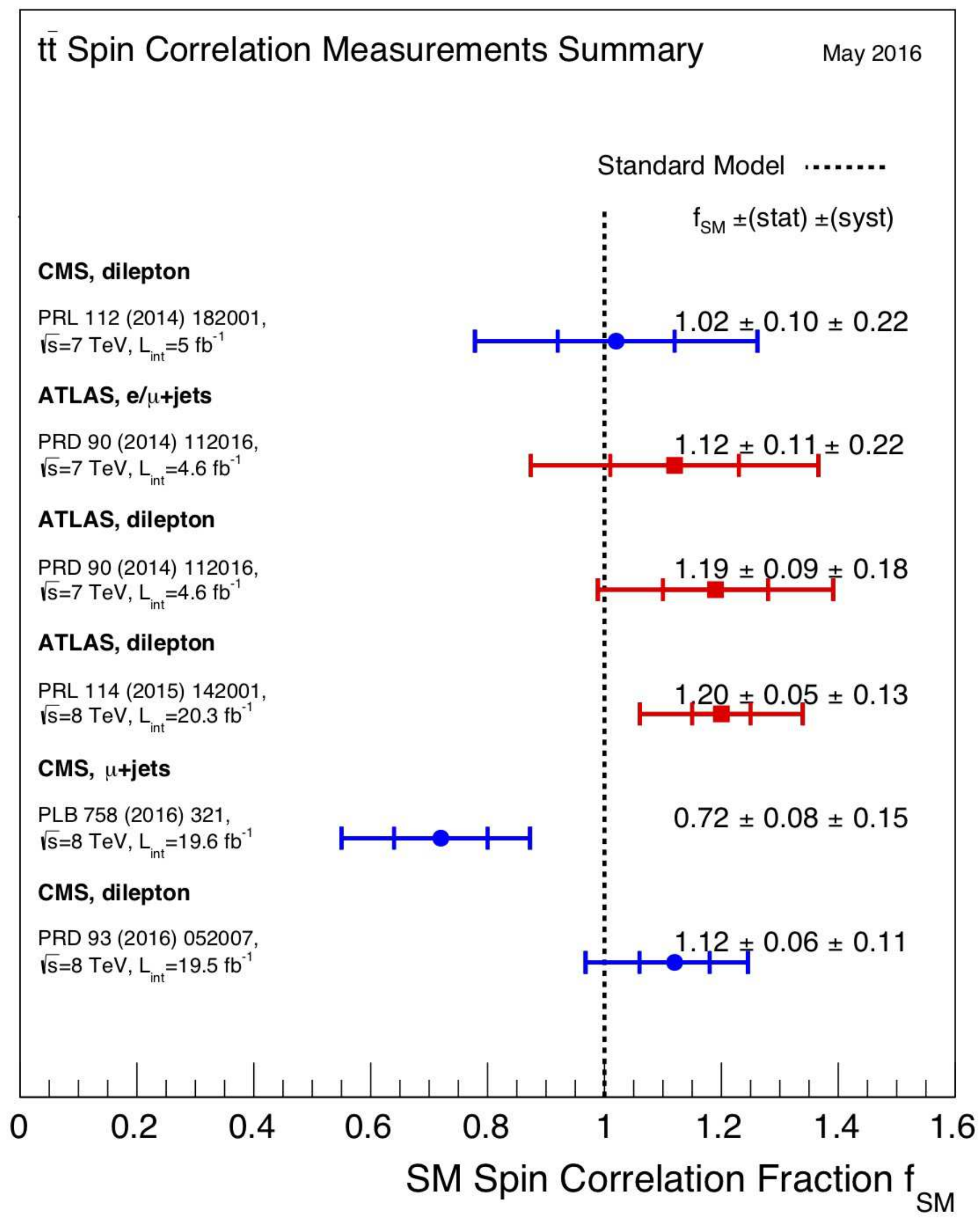}
\end{center}
\caption{\cite{reference:lhctopwg} Summary of $\rm t\bar{t}$ spin correlation measurements performed by ATLAS and CMS at
$\rm\sqrt{s}$ = 7 and 8 TeV using the LHC Run 1 datasets.}
\label{fig:spincorrelation_lhctop_comb}
\end{figure*}

CMS on the other hand performs differential measurements in bins of $\rm m_{t\bar{t}}$ to
reduce the systematic uncertainty associated to top quark $\rm p_T$ modeling, while exploring
two other observables, i.e., $\rm\cos\phi$ (the cosine of the opening angle between two lepton momenta
measured in the rest frames of their respective parent quarks, top or antitop), and
$\rm\cos\theta^*_{l^+}\cos\theta^*_{l^-}$ (the product of the cosines of the helicity angles of the leptons).
Overall, CMS observes a value of $\rm f_{SM}=1.12^{+0.12}_{-0.15}$, which is in good agreement with the SM
prediction. CMS translates the $\rm t\bar{t}$ spin correlation measurement to set exclusion limits on
the chromo-magnetic dipole moment. Here, the values outside the intervals 0.053$\rm <Re(\hat{\mu_t})<$0.026
and 0.068$\rm <Im(\hat{d_t})<$0.067 are excluded at 95\% CL.
All the $\rm t\bar{t}$ spin correlation measurements so far performed by the LHC experiments
are summarized in Fig.~\ref{fig:spincorrelation_lhctop_comb}. 
\section{Top Decay Width}
\begin{figure*}
\begin{center} 
\begin{tabular}{cc}
\includegraphics[width=0.45\textwidth]{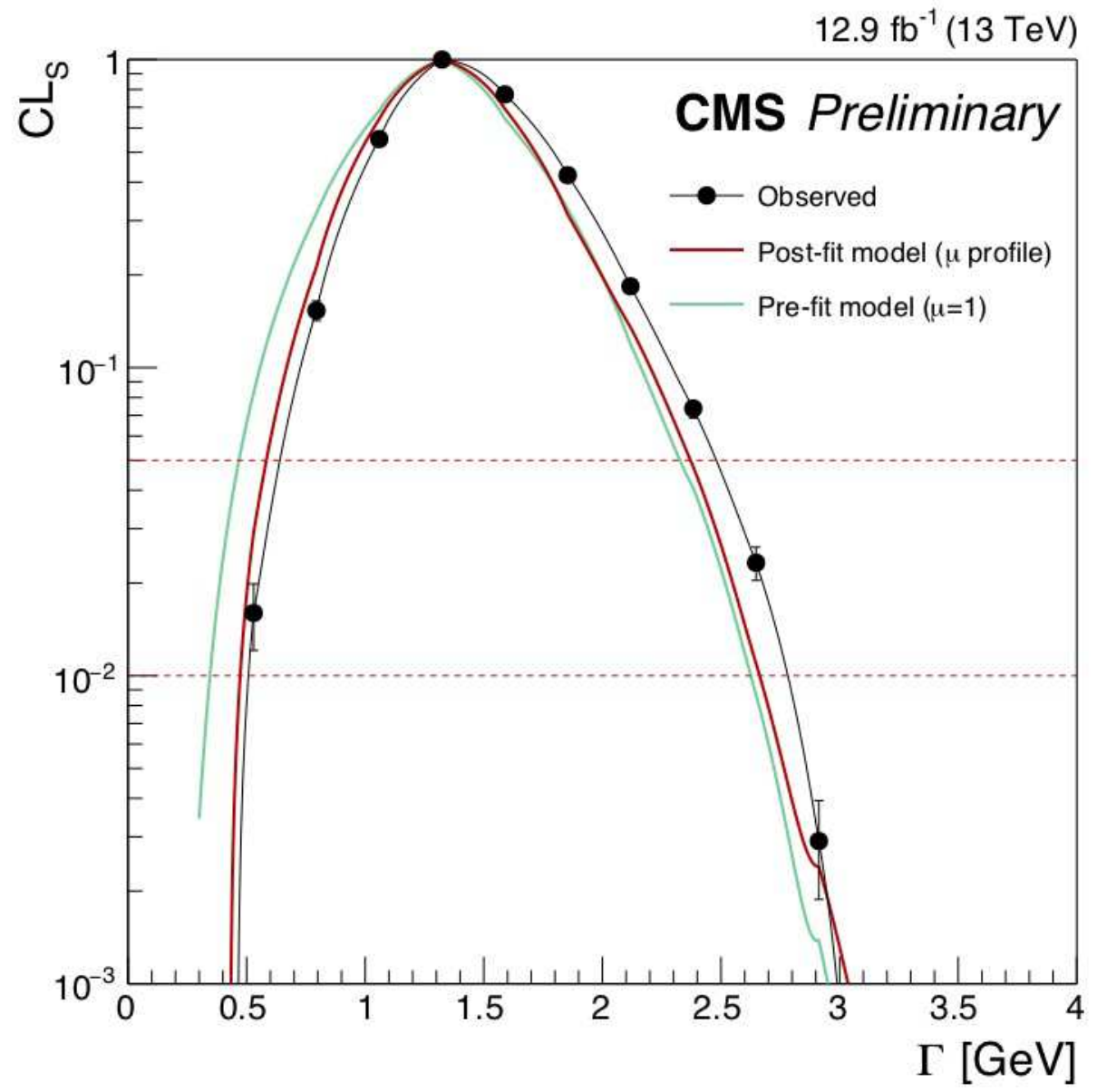} &\includegraphics[width=0.45\textwidth]{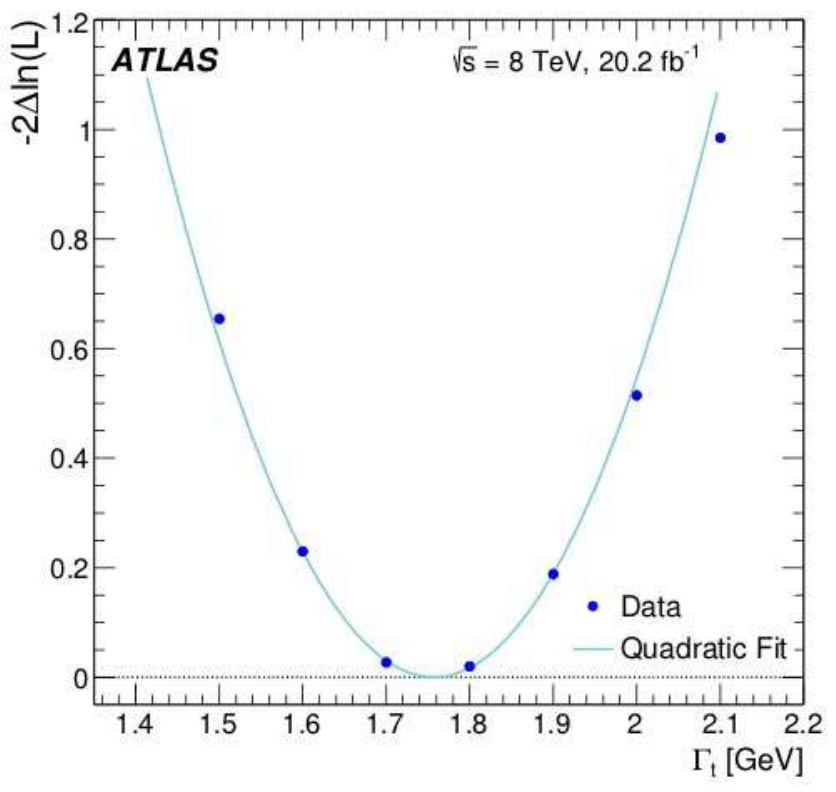}\\
(a) & (b)
\end{tabular}
\end{center}
\caption{\cite{reference:topwidth_cms}\cite{reference:topwidth_atlas} (a) Evolution of the $\rm CL_s$ as a function of the top quark width in CMS $\rm t\bar{t}\rightarrow$
dilepton events;
(b) Twice the negative log-likelihood distribution as a function of $\rm\Gamma_t$ in ATLAS $\rm t\bar{t}\rightarrow$
lepton+jet events.}
\label{fig:topwidth_cmsatlas}
\end{figure*}

Being quite heavy the top quark has a large decay width ($\rm\Gamma_t$). Within the SM,
the Next-to-next-to-leading-order (NNLO) calculations predict $\rm\Gamma_t$ of 1.322 GeV
for a top quark mass ($\rm m_{top}$) of 172.5 GeV and $\rm\alpha_s$=0.1189~\cite{reference:topwidth_theory}.
CMS has recently utilized the $\rm t\bar{t}\rightarrow$ dilepton events from $\rm 12.9 \  fb^{-1}$ of
the Run 2 dataset (at $\rm\sqrt{s}$ = 13 TeV) to constrain the total decay width of the top quark
through direct measurement. The analysis is based on a reconstructed observable, $\rm M_{lb}$,
i.e., the invariant mass of the lepton-jet (b-tagged) system. The constraint on $\rm\Gamma_t$
has been drawn through a two dimensional likelihood fit where the signal
strength ($\rm\mu=\sigma_{obs}/\sigma_{SM}$) and the sample fraction for an alternative width hypothesis
(denoted x) are varied. As shown in Fig.~\ref{fig:topwidth_cmsatlas}(a), the likelihood fit
provides an observed (expected) bound of $\rm 0.6 <\Gamma_t < 2.5\  (0.6<\Gamma_t<2.4)$ GeV
at 95\% confidence level~\cite{reference:topwidth_cms}.
\par
ATLAS performed a more refined measurement of top quark decay width using the $\rm t\bar{t}\rightarrow$
lepton+jets events from $\rm 20.2 \  fb^{-1}$ of the Run 1 dataset at $\rm\sqrt{s}$ = 8 TeV. Here, the template fitting
method uses various kinematic observables associated with the hadronically and semileptonically
decaying top quarks. As shown in Fig.~\ref{fig:topwidth_cmsatlas}(b), the measurement yields a value of 
$\rm\Gamma_t=1.76\pm 0.33 (stat)^{+0.79}_{-0.68}$ (syst) GeV (for $\rm m_{top}$=172.5 GeV)
~\cite{reference:topwidth_atlas},
in good agreement with the SM predicted value. However, the measurement is limited by
the systematic uncertainties from jet energy scale/resolution and signal modeling.

\section{W-boson Helicity}
\begin{figure*}
\begin{center}
\includegraphics[width=0.9\textwidth]{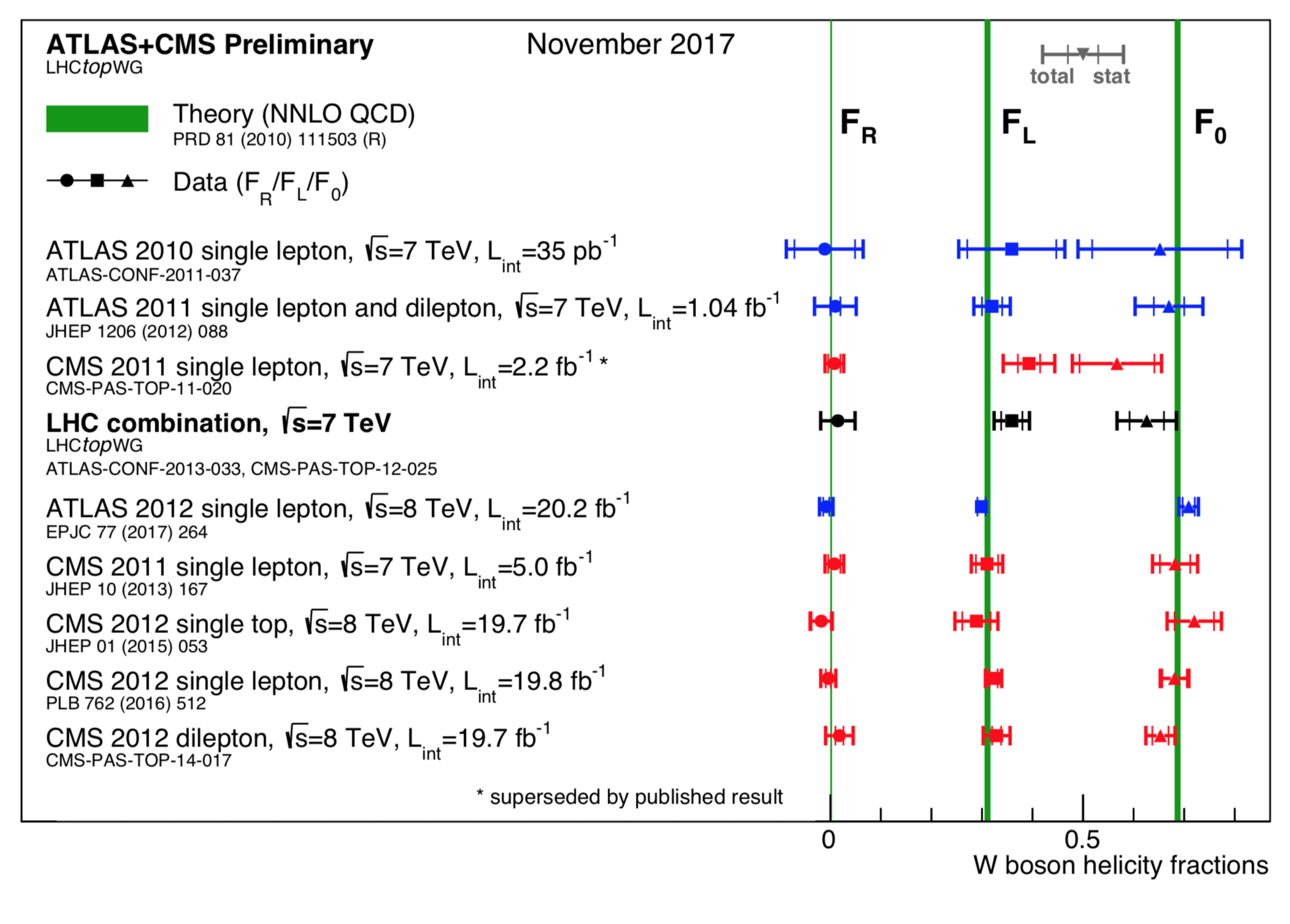}
\end{center}
\caption{\cite{reference:lhctopwg} Measured values of W-boson helicity fractions, $\rm F_{L,R,0}$ from ATLAS and CMS
during LHC Run 1.}
\label{fig:whelicity}
\end{figure*}

Within the SM the top quarks decays exclusively into a W-boson and a b-quark
and the weak Wtb vertex has a (V-A) structure, where V and A respectively refer
to the vector and axial vector components of the weak coupling. As the W-boson
appears as a real particle in the top quark decays, its polarization can be measured
using the $\rm t\bar{t}$ events. Here, the W-boson helicity fractions (left-handed,
right-handed and longitudinal) are defined as $\rm F_{L,R,0} = \Gamma_{L,R,0}/\Gamma_{Total}$,
where $\rm \Gamma_{L,R,0}$ are the partial decay widths in left-handed, right-handed, and
longitudinal helicity states respectively, with $\rm \Gamma_{Total}$ being the
total decay width. The SM next-to-next-to-leading order (NNLO) calculations
\cite{reference:whelicity_sm} including the electroweak effects predict the values of
$\rm F_L = 0.311\pm 0.005$, $\rm F_R = 0.0017\pm 0.0001$ and $\rm F_0 = 0.687\pm 0.005$,
for a top quark mass of $\rm 172.8\pm 1.3$ GeV. Experimentally, the helicity angle
$\rm\theta^*$ can be defined as the angle between the direction of either the
down-type quark or the charged lepton arising from the W-boson decay and the reversed
direction of the top quark, both in the rest frame of the W-boson. The differential
cross-sections as a function of $\rm\cos\theta^*$ can then be written as
\begin{eqnarray*}
\rm \frac{1}{\sigma}\frac{d\sigma}{d\cos\theta^*} = \frac{3}{8} (1-\cos\theta^*)^2 F_L 
+\frac{3}{8} (1+\cos\theta^*)^2 F_R
+\frac{3}{4} (\sin\theta^*)^2 F_0.
\end{eqnarray*}
Both ATLAS and CMS have performed the measurements using $\rm t\bar{t}\rightarrow$ lepton+jets
events based on $\rm 20.2 \  fb^{-1}$ and $\rm 19.8\  fb^{-1}$ datasets respectively from 2012 LHC
operations. The observable related to the W-boson helicity fractions, $\rm \cos\theta^*$,
is reconstructed considering the leptonic ($\rm t\rightarrow bW\rightarrow bl\nu$) and hadronic
($\rm t\rightarrow bW\rightarrow bq\bar{q}'$) decay branches of the top quark.
The CMS measurements~\cite{reference:whelicity_cms}
result in $\rm F_L = 0.323\pm 0.008 (stat)\pm 0.014 (syst)$,
$\rm F_R = 0.004\pm 0.005 (stat)\pm 0.014 (syst)$ and 
$\rm F_0 = 0.681\pm 0.012 (stat)\pm 0.023 (syst)$, while the ATLAS observes
\cite{reference:whelicity_atlas} the values of
$\rm F_L = 0.299\pm 0.015$, $\rm F_R = 0.008\pm 0.014$ and
$\rm F_0 = 0.709\pm 0.019$. 
Fig.~\ref{fig:whelicity} shows all the measurements from ATLAS and CMS during the LHC Run 1, along
with the combined measurements (at $\rm\sqrt{s}$ = 7 TeV).

\section{Top Quark Mass}
The top quark mass, $\rm m_{top}$ is a key parameter in the SM and is the major contributor
to the Higgs boson mass ($\rm m_H$) through radiative corrections. Therefore, the accuracy
on both $\rm m_{top}$ and $\rm m_H$ measurements is quite crucial for the consistency tests of the SM.
Starting from the Tevatron experiments, the top quark mass has so far been measured with
increasing precision using multiple final states, as well as with different analysis techniques.
Two of the most recent $\rm m_{top}$ measurements from CMS and ATLAS experiments are
presented here.
\par
\begin{figure*}
\begin{center}
\begin{tabular}{cc}
\includegraphics[width=0.45\textwidth]{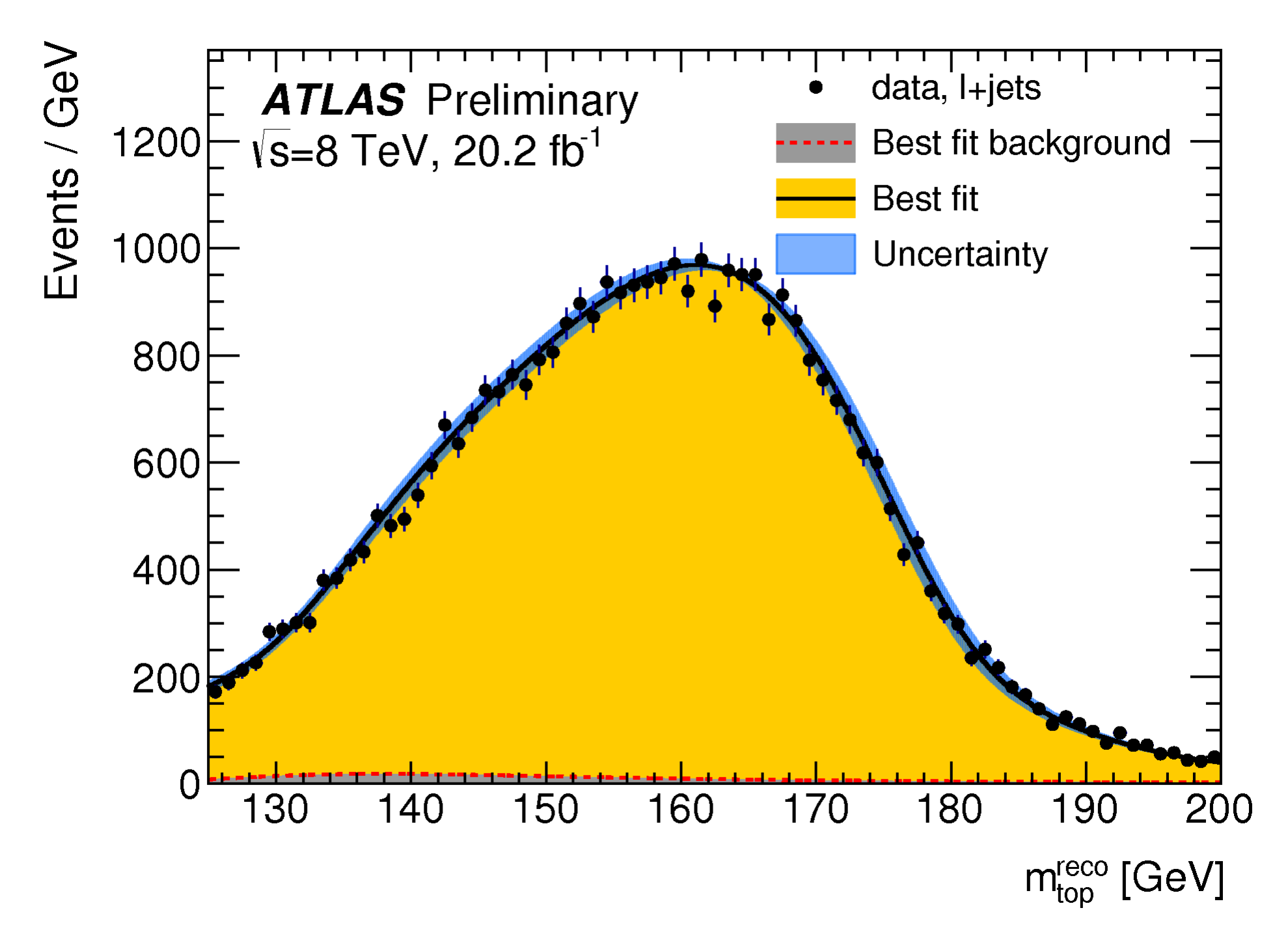}&\includegraphics[width=0.45\textwidth]{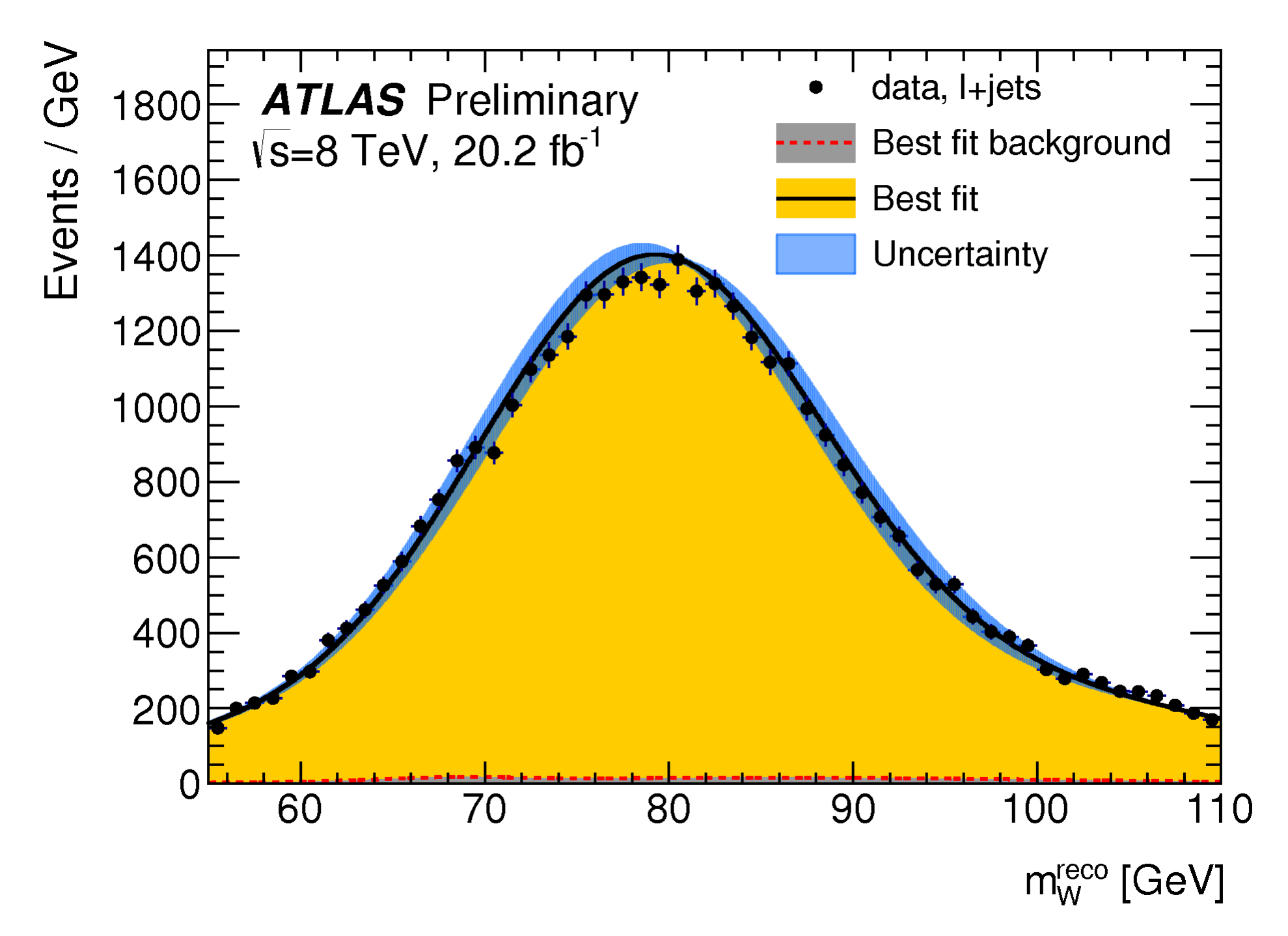}\\
(a) & (b)
\end{tabular}
\end{center}
\caption{\cite{reference:mtop_atlas} After the template fit, comparison between data, signal and
background for
(a) reconstructed top mass and (b) reconstructed W-boson mass.}
\label{fig:mtop_atlas}
\end{figure*}

ATLAS has recently performed a top quark mass measurement~\cite{reference:mtop_atlas}
in lepton+jets final states using a $\rm 20.2\  fb^{-1}$ dataset at $\rm\sqrt{s}$ = 8 TeV. The full event
reconstruction is performed using a likelihood based kinematic fitter, KLFITTER~\cite{reference:mtop_klfitter}.
The $\rm t\bar{t}\rightarrow$ lepton+jets event selection is further optimized through the usage
of a boosted decision tree~\cite{reference:tmva}. The top quark mass ($\rm m_{top}$) together with the jet energy scale
factor (JSF) and b-jet energy scale factor (bJSF) is then simultaneously extracted using the template
fit technique. The template fit
results in terms of $\rm m_{top}$ and $\rm m_W$ are shown in Fig.~\ref{fig:mtop_atlas}. The measurement
yields a top quark mass of $\rm 172.8\pm 0.39 (stat)\pm 0.82 (syst)$ GeV, where the dominant uncertainties
are driven by theoretical modeling and systematics.
\par
\begin{figure*}
\begin{center}
\begin{tabular}{cc}
\includegraphics[width=0.45\textwidth]{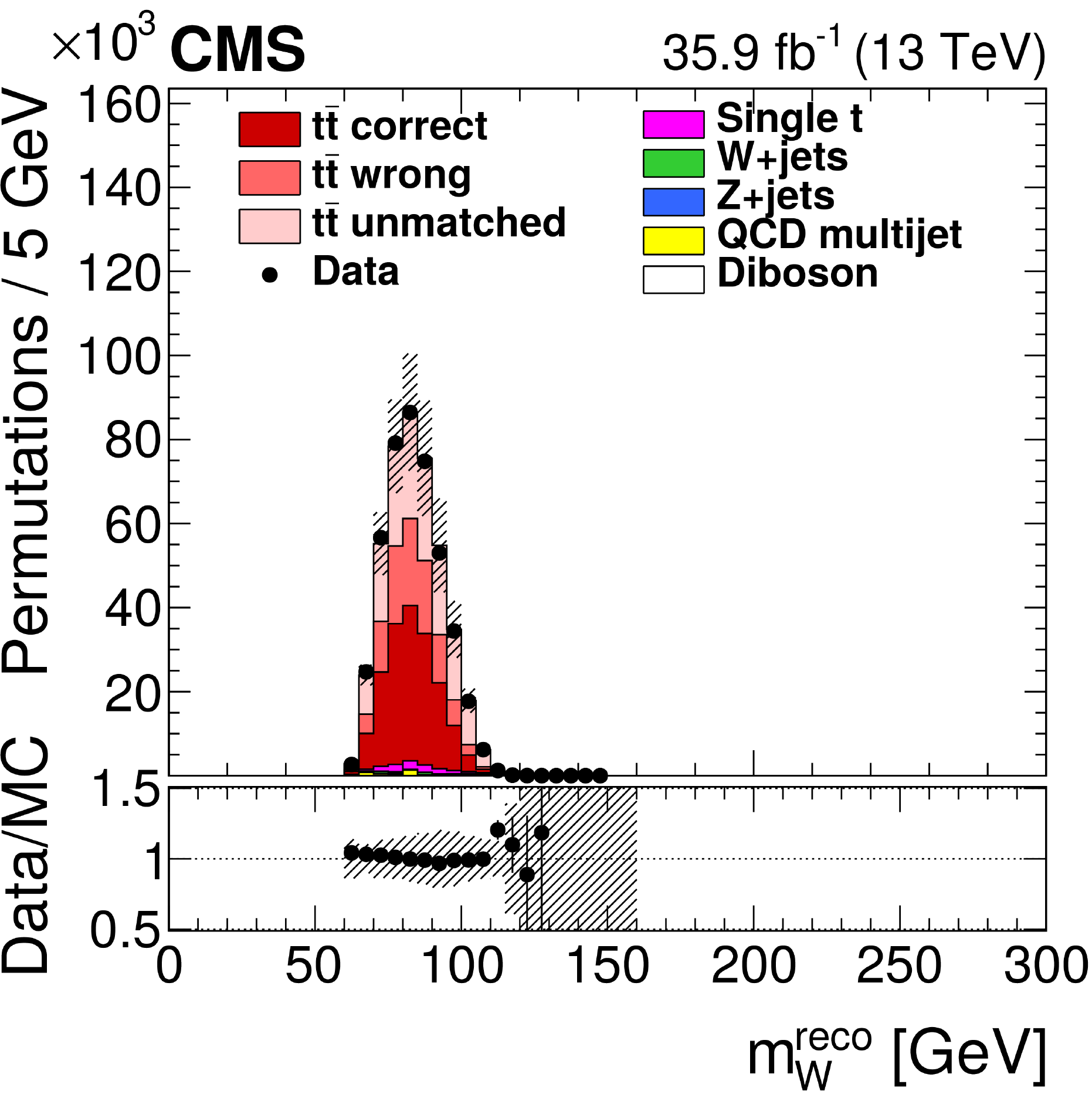}&\includegraphics[width=0.45\textwidth]{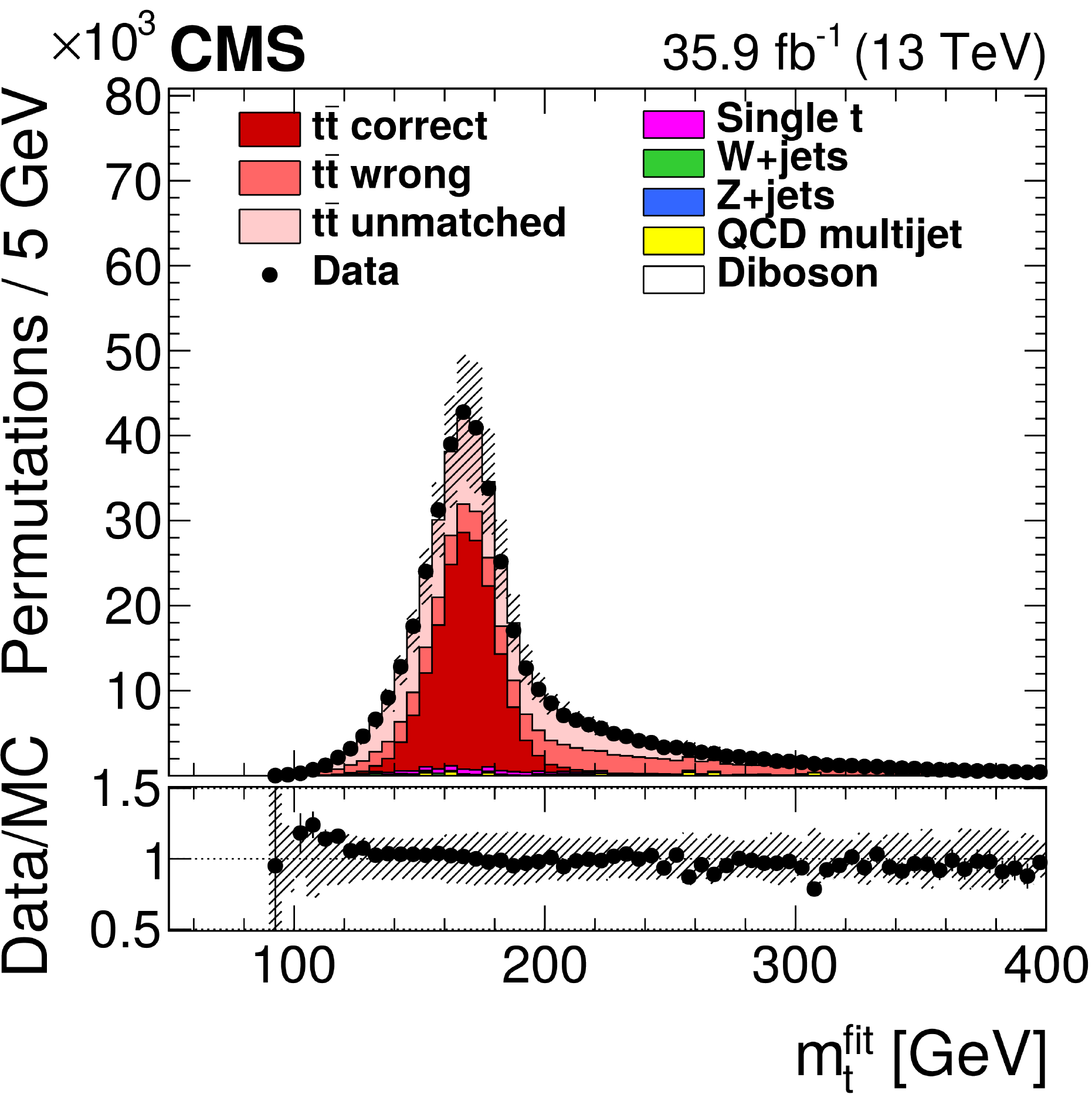}\\
(a) & (b)
\end{tabular}
\end{center}
\caption{\cite{reference:mtop_cms} Distribution of (a) reconstructed $\rm m_W$, and (b) $\rm m_{top}$ for all selected permutations.}
\label{fig:mtop_cms}
\end{figure*}
\begin{figure*}
\begin{center}
\includegraphics[width=0.85\textwidth]{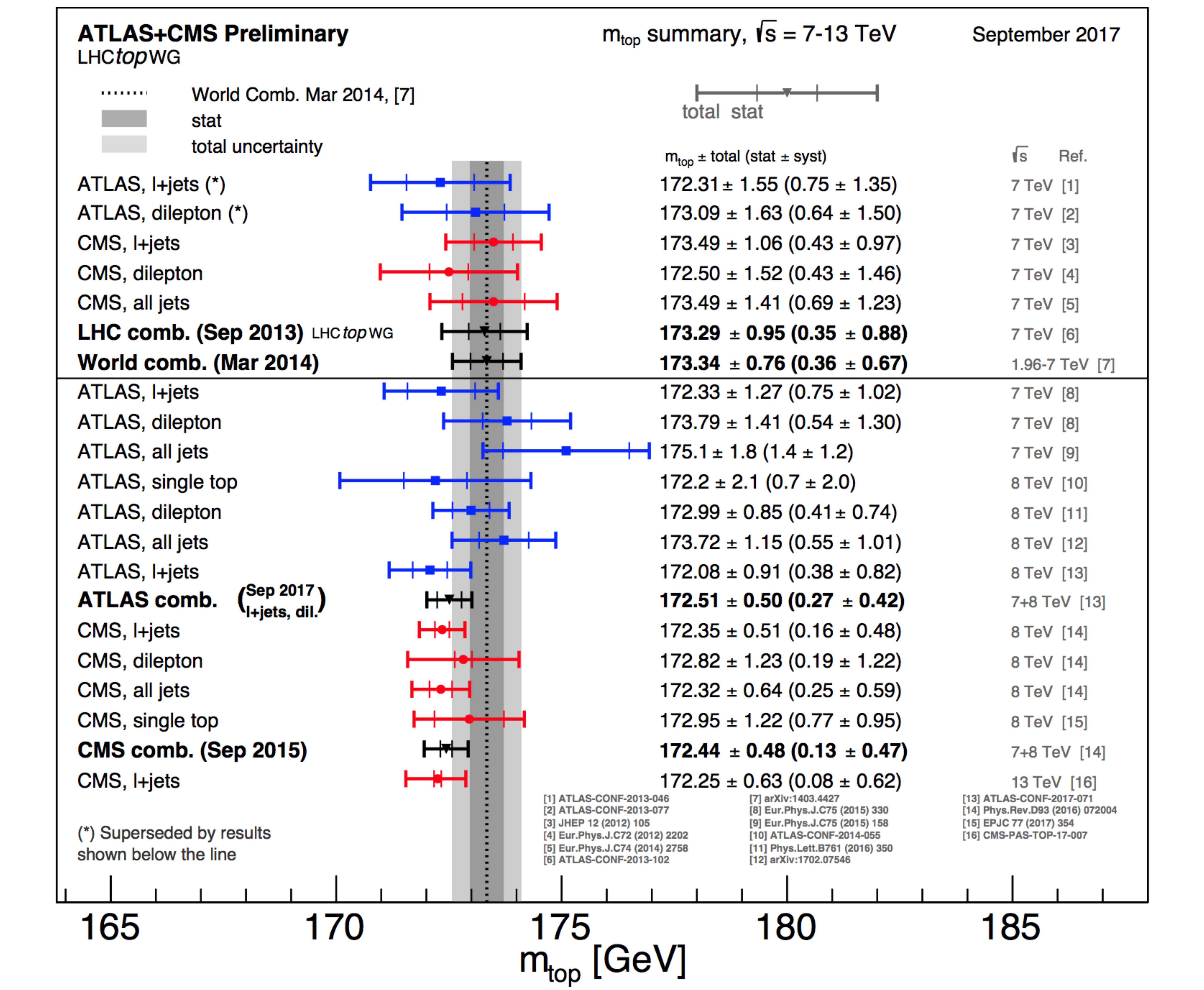}
\end{center}
\caption{\cite{reference:lhctopwg} ATLAS and CMS $\rm m_{top}$ results (individual and combined), along with the world average.}
\label{fig:mtop_lhctop}
\end{figure*}

The latest $\rm m_{top}$ measurement~\cite{reference:mtop_cms} from CMS is based on a $\rm 35.9 \  fb^{-1}$
dataset at $\rm\sqrt{s}$ = 13 TeV. The full $\rm t\bar{t}\rightarrow$ lepton+jets event reconstruction is
performed using a kinematic fit of the decay products.
A 2-D ideogram fitting technique~\cite{reference:mtop_ideogram} is then applied to
the data to measure the top quark mass simultaneously with an overall jet energy scale factor (JSF),
constrained by $\rm m_W$ (through $\rm W\rightarrow q\bar{q}^\prime$ decays); the fit results in terms
of $\rm m_{top}$ and $\rm m_W$
are shown in Fig.~\ref{fig:mtop_cms}. The ideogram method measures an $\rm m_{top}$ value
of $\rm 172.25\pm 0.08 (stat)\pm 0.62 (syst)$ GeV, in consistency with the Run 1 CMS measurements at
$\rm\sqrt{s}$ = 7 and 8 TeV. The measurement results in a precision of $\rm\Delta m_{top}/m_{top}\approx$
0.36\% where the leading uncertainties originate from MC modeling, color reconnection, parton showering, JES, etc.
The most recent individual $\rm m_{top}$ measurements from the LHC experiments, along with the world
average value for $\rm m_{top}$ are summarized in Fig.~\ref{fig:mtop_lhctop}.

\section{Conclusions}
Some of the key measurements related to top quark properties from the ATLAS and CMS experiments
have been presented here using $\rm t\bar{t}$ pair production. The measurements are at large
consistent with SM predictions and with enhanced statistical precision. The systematic
uncertainties of these measurements are limited by the theoretical modeling (top mass,
$\rm t\bar{t}$ scale, $\rm t\bar{t}$ matching scale, etc.). In addition, experimental
uncertainties related to the jets (Jet Energy Scale, Jet Energy Resolution) affect the
measurement precision. Further details about these measurements can be found
in ATLAS and CMS web pages~\cite{reference:atlas_top_public results}\cite{reference:cms_top_public results}
collecting public results.

\end{document}